\documentclass[apj]{emulateapj}
\usepackage[colorlinks,linkcolor={blue},citecolor={blue},urlcolor={red}]{hyperref}
\usepackage{epsfig}
\usepackage{graphicx}
\usepackage{natbib}
\usepackage{amsmath}
\usepackage{amsfonts}
\usepackage{amssymb}

\newcommand{\myemail}{\email{leech@shao.ac.cn}}

\shorttitle{Linking galaxies to halos with stellar mass or with stellar velocity dispersion?}
\shortauthors{Li, Wang, Jing}

\begin{document}

\title{Stellar mass versus stellar velocity dispersion: which is 
better for linking galaxies to their dark matter halos?}

\author{Cheng  Li\altaffilmark{1}, Lixin Wang\altaffilmark{1},
and Y. P.  Jing\altaffilmark{2}}
\myemail

\altaffiltext{1}{Partner   Group   of  the   Max   Planck  Institute   for
  Astrophysics  at  the  Shanghai  Astronomical  Observatory  and  Key
  Laboratory for Research in Galaxies and Cosmology of Chinese Academy
  of   Sciences,    Nandan   Road   80,    Shanghai   200030,   China}
\altaffiltext{2}{Center for Astronomy and Astrophysics, Shanghai Jiao Tong 
University, Shanghai 200240, China}

\begin{abstract}
It  was  recently  suggested   that,  compared  to  its  stellar  mass
($M_\ast$), the central stellar velocity dispersion ($\sigma_\ast$) of
a galaxy  might be a  better indicator for  its host dark  matter halo
mass.  Here we test this hypothesis by estimating the dark matter halo
mass  for central galaxies  in groups  as a  function of  $M_\ast$ and
$\sigma_\ast$.   For   this  we  have   estimated  the  redshift-space
cross-correlation function (CCF) between the central galaxies at given
$M_\ast$ and  $\sigma_\ast$ and a reference galaxy  sample, from which
we  determine both  the projected  CCF, $w_p(r_p)$,  and  the velocity
dispersion  profile (VDP).   A halo  mass  is then  obtained from  the
average  velocity  dispersion  within  the virial  radius.   At  fixed
$M_\ast$, we  find very weak or  no correlation between  halo mass and
$\sigma_\ast$.  In  contrast, strong  mass dependence is  clearly seen
even when $\sigma_\ast$ is limited to a narrow range. Our results thus
firmly demonstrate that the stellar  mass of central galaxies is still
a good (if  not the best) indicator for dark  matter halo mass, better
than  the  stellar  velocity  dispersion.  The  dependence  of  galaxy
clustering on $\sigma_\ast$ at  fixed $M_\ast$, as recently discovered
by  \citet{Wake-Franx-vanDokkum-12}, may  be  attributed to  satellite
galaxies,  for which  the tidal  stripping occurring  within  halos has
stronger effect on stellar mass than on central stellar velocity dispersion.
\end{abstract}


\section{Introduction}
\label{sec:introduction}

The stellar mass of central  galaxies in dark matter halos is believed
to be  strongly correlated with the  dark matter mass  of their halos.
This relationship has been extensively studied in recent years using a
variety of observational probes including galaxy clustering, satellite
kinematics, gravitational  lensing and group/cluster  catalogs. It has
also  formed  the  basis  for   most  (if  not  all)  of  the  current
physical/statistical models  that aim at populating  dark matter halos
with galaxies,  such as  semi-analytic models (SAMs),  halo occupation
distribution  (HOD)  models  and  subhalo  abundance  matching  models
(SHAMs).   These  studies  have  well  established  that  the  stellar
mass--halo mass  relation for central  galaxies can be described  by a
double power-law form  with a relatively small scatter  of $\sim 0.16$
dex \citep[e.g.][]{vandenBosch-04, Wang-06, Wang-07, Yang-07, More-09,
  Yang-Mo-vandenBosch-09,    Behroozi-Conroy-Wechsler-10,   Moster-10,
  More-11, Guo-10, Li-12a, Yang-12}.

Using data from the Sloan Digital Sky Survey \citep[SDSS;][]{York-00},
\citet{Wake-Franx-vanDokkum-12} recently showed that, when compared to
stellar  mass  ($M_\ast$),  the  central stellar  velocity  dispersion
($\sigma_\ast$)  of   galaxies  is  more  closely   related  to  their
clustering  properties.    This  led  the  authors   to  suggest  that
$\sigma_\ast$  might  be  better  than $M_\ast$  when  indicating  the
properties  of dark matter  halos that  determine clustering,  such as
halo  mass or assembly  history.  On  the other  hand, as  the authors
pointed out,  their finding cannot  rule out the possibility  that the
correlation of dark  matter halos with $M_\ast$ is  still tighter than
that with $\sigma_\ast$.  It is known that satellite galaxies may well
deviate from the stellar mass--halo mass relation of central galaxies,
due to the stripping of their outer regions by tidal interactions with
their  host halos,  which  has  stronger effect  on  $M_\ast$ than  on
$\sigma_\ast$.

This motivates our  work, in which we attempt  to discriminate between
these possibilities  by directly measuring  the dark matter  halo mass
for central galaxies of  different stellar masses and stellar velocity
dispersions. For this we first estimate the cross-correlation function
(CCF) in redshift space between a reference sample of galaxies and the
central galaxies  of groups with given $M_\ast$  and $\sigma_\ast$. We
then estimate  the velocity  dispersion profile of  satellite galaxies
around the  central galaxies by  modelling the redshift  distortion in
the CCF, from  which we determine an average mass  for the dark matter
halos  in  which the  central  galaxies  reside.   In a  recent  paper
\citep[][hereafter  Paper  I]{Li-12a}, we  have  shown  that 
for  central galaxies with different  luminosities and  masses,
the  dark matter  halo masses  measured  in  this way  
are in  good  agreement with  the results obtained by \citet{Mandelbaum-06} 
from weak lensing analysis of the SDSS data.
When compared to
the       galaxy-galaxy       cross-correlations       probed       in
\citet{Wake-Franx-vanDokkum-12},    the    cross-correlation   between
galaxies and group  central galaxies enables us to  directly probe the
central  galaxy --  halo mass  relation, thus  avoiding the  effect of
satellite galaxies.  In addition,  the velocity dispersion of satellite
galaxies is  caused by the local  gravitational field, thus  is a more
direct measure of dark matter  halo mass than the clustering amplitude
adopted in \citet{Wake-Franx-vanDokkum-12}.

Throughout we assume a $\Lambda$ cold dark matter cosmology model with
$\Omega_m=0.27$, $\Omega_\Lambda=0.73$ and $h=0.7$.

\begin{figure}
  \begin{center}
    \epsfig{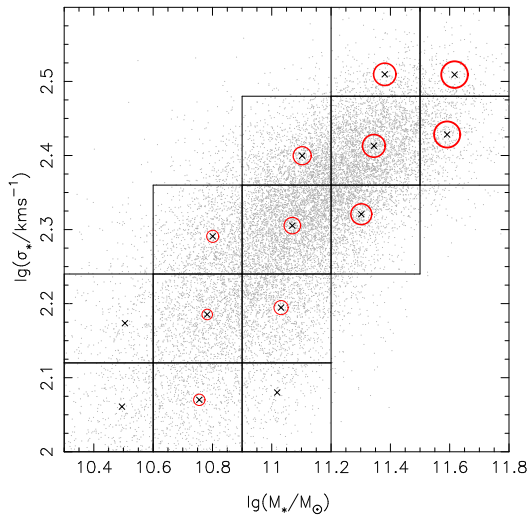}
  \end{center}
    \caption{Distribution  of the  central  galaxies in  the SDSS  DR7
      galaxy group catalog in the plane of stellar mass ($M_\ast$) and
      stellar velocity dispersion ($\sigma_\ast$). The samples used
      in  this work  are  indicated  by the  boxes,  with the  average
      $M_\ast$  and $\sigma_\ast$  being  marked by  a  cross in  each
      box. The estimated halo mass  for each sample is indicated by
      the size of the red circles which is scaled by the halo mass.}
  \label{fig:samples}
\end{figure}

\section{Data and methodology}
\label{sec:data}

We  apply   our  analysis  to   the  SDSS  galaxy  group   catalog  of
\citet{Yang-07}.  This  is  a  catalog  of local  galaxy  groups  with
$0.01<z<0.2$, and  is constructed  from {\tt sample  dr72} of  the New
York        University        Value-Added        Galaxy        Catalog
\citep[NYU-VAGC;][]{Blanton-05}  using   a  halo-based  group  finding
algorithm  developed in  \citet{Yang-05a}.  The  stellar mass  of each
galaxy, $M_\ast$, accompanies the NYU-VAGC release, which is estimated
by  \citet{Blanton-Roweis-07} from the  SDSS redshift  and photometric
data,  assuming   a  universal   stellar  initial  mass   function  of
\citet{Chabrier-03}.   We  use the  ``total  masses''  instead of  the
``Petrosian  masses'', obtained  by correcting  the latter  using SDSS
``model  magnitudes'' (see \citealt{Li-White-09}  and \citealt{Guo-10}
for details). The  central stellar velocity dispersion ($\sigma_\ast$)
is  available for  each galaxy  from the  SDSS  spectroscopy, measured
within  the  3$^{\prime\prime}$  diameter  fiber.  We  have  corrected
$\sigma_\ast$ to  an aperture  of one eighth  of the  galaxy effective
radius           following          \citet{Cappellari-06}          and
\citet{Wake-Franx-vanDokkum-12}.

As in Paper I, we define  the most massive galaxy in each group
as its central galaxy. It has been suggested that brightest cluster 
galaxies (BCGs), or the most massive galaxies adopted here, may be not 
exactly located at their halo center. However, a recent study by 
\cite{vonderLinden-12} showed that such offsets are small in general; 
the average offset between X-ray centroids and BCGs are $\sim20$ kpc.
We  use all  the central galaxies of which the  host groups
have three or more member galaxies.  This leads to a number of $\sim
16,000$  central galaxies.  We  divide the  central  galaxies into  14
samples  according   to  their  stellar  mass   and  central  velocity
dispersion.  In  Figure~\ref{fig:samples} we plot  the distribution of
all the central  galaxies in our samples in the  plane of $M_\ast$ and
$\sigma_\ast$.  The regions  of the  different samples are  indicated  by the
boxes, while the  average mass and velocity dispersion  of each sample
are marked with a cross.  This  selection scheme gives us at least two
$\sigma_\ast$ ($M_\ast$) samples  at a given $M_\ast$ ($\sigma_\ast$),
enabling us to study the dependence of clustering and halo mass on one
of the  two parameters  while fixing the  other. Our  samples cover
similar  $M_\ast$  and $\sigma_\ast$  ranges  as  those considered  in
\citet{Wake-Franx-vanDokkum-12}.

For each of the 14 central  galaxy samples, we begin by estimating the
redshift-space cross-correlation function (CCF), $\xi^{(s)}(r_p,\pi)$,
with respect to  a reference galaxy sample selected  from the NYU-VAGC
{\tt sample  dr72}. 
The reference sample consists of about half a million galaxies with
$r$-band apparent magnitudes $r<17.6$ and absolute magnitudes 
$-24<M_{^{0.1}r}<-16$, and redshifts in the range $0.01<z<0.2$. Details 
of the sample selection can be found in Paper I.
We then  obtain the projected CCF,  $w_p(r_p)$, by
integrating  $\xi^{(s)}(r_p,\pi)$ over  the line  of  sight separation
$\pi$.   Next,  we model the real-space  CCF $\xi_{cg}(r)$ with a 
combination of an  \citet{Navarro-Frenk-White-97} profile and a biased 
linear autocorrelation of dark matter, and we determine an accurate  
description of $\xi_{cg}(r)$ by fitting the
Abel  transform  of  the   model  to  the  observed  $w_p(r_p)$.   The
one-dimensional velocity  dispersion profile (VDP)  of galaxies around
the  central  galaxies in  each  sample  is  then estimated by 
comparing $\xi^{(s)}(r_p,\pi)$  with $\xi_{cg}(r)$, through modeling
the redshift  distortion   in  $\xi^{(s)}(r_p,\pi)$.
Finally,  we use  $N$-body cosmological simulations 
to  calibrate the relationship between  the so-obtained velocity  
dispersion (VD) and  the  dark matter  halo mass,
with which we determine a halo  mass for each  of our central  galaxy 
sample.  Details  of our
methodology, as well as the  reference sample and simulations used for
the computation and calibration can be  found in Paper I.  In this work
we  estimate  errors  on  all  the measurements  using  the  bootstrap
resampling technique \citep{Barrow-Bhavsar-Sonoda-84}.

\section{Results}

\begin{figure}[t]
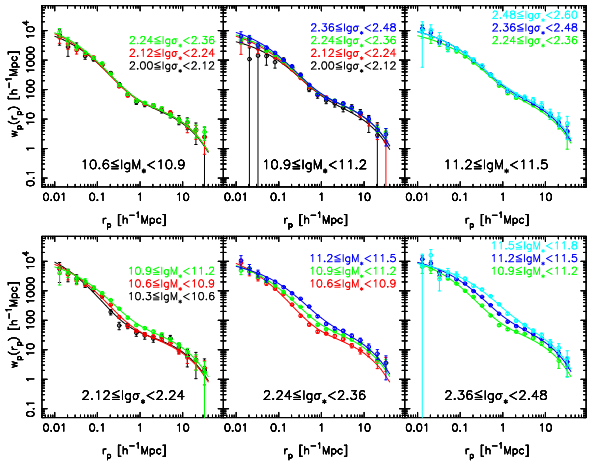

  \begin{center}
    \epsfig{figure=f2a.ps,width=0.9\hsize}
    \epsfig{figure=f2b.ps,width=0.9\hsize}
  \end{center}
    \caption{The  projected  cross-correlation  function  between  the
      central  galaxies  of groups  and  the  reference galaxies,  for
      central  galaxies  of different  stellar  masses ($M_\ast$)  and
      stellar  velocity dispersions  ($\sigma_\ast$), as  indicated in
      each panel. Results  from the SDSS DR7 galaxy  group catalog are
      plotted in colorful symbols  with error bars, and the solid lines
      are best-fit models (see the text for details).}
  \label{fig:wrp}
\end{figure}

In Figure~\ref{fig:wrp}, we show  the projected CCFs determined in the
way described above  for some of our samples.  We plot  the results for samples
of different $\sigma_\ast$ but fixed $M_\ast$ in the upper panels, and
the results for samples  of different $M_\ast$ but fixed $\sigma_\ast$
in the lower panels. At fixed stellar mass, the projected CCF shows no
or very weak dependence on $\sigma_\ast$, and this is true for all the
masses considered  ($10.3<\lg(M_\ast/M_\odot)<11.8$) and at  all scales
probed   ($\sim$15  kpc$<r_p<$ $\sim$30  Mpc).   In   contrast,  the
projected CCF at fixed  $\sigma_\ast$ shows significant and systematic
trends with  mass, in both amplitude  and slope, and this  is true for
all  the  $\sigma_\ast$  bins.    The  CCF  amplitude  increases  with
increasing  mass at  all  scales above  $\sim  100$ kpc.  This
reflects the tight correlation between the stellar mass of the central
galaxies and their dark matter halo mass, which clearly holds
even when  the central stellar  velocity dispersion of the  galaxies is
limited to a narrow range.  Moreover, the one-halo term 
below $\sim$ 1 Mpc shows steeper slopes  at lower  masses and  flatter 
slopes at higher  masses, implying more centrally concentrated distribution
of satellite galaxies in less-massive halos (see Paper I for detailed
discussion).

\begin{figure}[t]
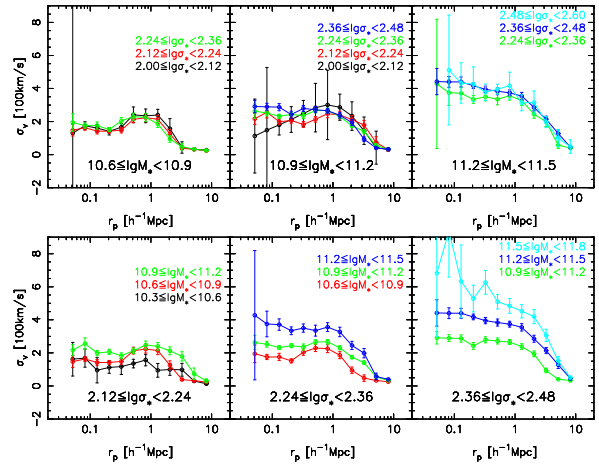

  \begin{center}
    \epsfig{figure=f3a.ps,width=0.9\hsize}
    \epsfig{figure=f3b.ps,width=0.9\hsize}
  \end{center}
    \caption{Velocity dispersion profile measured for the SDSS DR7
    galaxy groups with central galaxies of different stellar masses ($M_\ast$)
    and stellar velocity dispersion ($\sigma_\ast$), as indicated
    in each panel.}
  \label{fig:vdp}
\end{figure}

In Figure~\ref{fig:vdp}, we show the velocity dispersion profile (VDP)
of  satellite  galaxies   around  our  central  galaxies,  $\sigma_v$,
measured as a function of the projected separation $r_p$ for different
$M_\ast$ and $\sigma_\ast$ samples. Similarly, we see significant mass
dependence at fixed  $\sigma_\ast$, but very weak or  no dependence of
the VDP on  $\sigma_\ast$ at a given mass.  We would like to point out
two interesting trends that can be read from the lower panels of the figure.
First,  at scales  smaller  than $\sim  1$  Mpc, the  velocity
dispersion increases with increasing central galaxy mass, with
more remarkable effect at higher masses.  The  velocity
dispersion of  galaxies in a group  or cluster is caused  by the local
gravitational field  and so, when compared to  the large-scale amplitude
of CCFs, it provides a more direct and reliable measure of the mass of
the host  dark matter halo.   Thus, the dependence of  the small-scale
$\sigma_v$  on $M_\ast$ at  fixed $\sigma_\ast$  shows again  that the
stellar mass of central galaxies  is more tightly correlated with halo
mass than  their central stellar  velocity dispersion is.   Second, at
lower masses  ($\la 10^{11} M_\odot$), a mass-dependent  shift is seen
in the {\em transition scale}  where the velocity dispersion starts to
deviate from  a flat  slope and increase  to form  a bump at  around 1
Mpc. As  shown in  Paper I (see  their fig.4),  the transition
occurs  at around  the virial  radius of  the host  dark  matter halo.
Therefore, the fact  that the transition is seen  at larger scales for
higher  stellar  masses  at  fixed $\sigma_\ast$  reflects  the  tight
correlation between the  central galaxy mass and the  virial radius of
its host dark matter halo, and thus the halo mass.

\begin{figure*}[t]
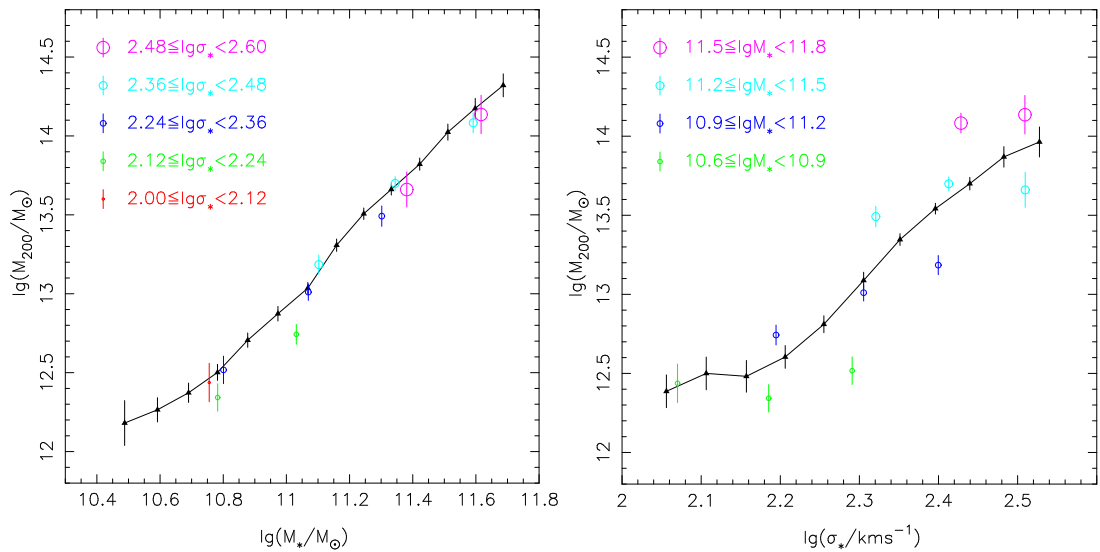

  \begin{center}
    \epsfig{figure=f4a.ps,height=0.4\hsize}
    \epsfig{figure=f4b.ps,height=0.4\hsize}
  \end{center}
  \vspace{-0.3cm}
    \caption{Dark matter halo mass  as function of galaxy stellar mass
      ($M_\ast$,  left-hand  panel)  and stellar  velocity  dispersion
      ($\sigma_\ast$,  right-hand  panel),  measured for  the  central
      galaxies of groups in the SDSS DR7. The colorful symbols are for
      the  samples selected  on the  $M_\ast$  versus $\sigma_\ast$
      plane as shown in Figure~\ref{fig:samples}.  The solid triangles
      connected with the line in each panel is for samples selected
      only by $M_\ast$ (left panel) or $\sigma_\ast$ (right panel).}
  \label{fig:mhalo}
\end{figure*}

We have estimated a halo mass  for each central galaxy sample based on
a relation between the velocity dispersion measured in our methodology
and    the    dark    matter    halo    mass,    as    described    in
$\S$~\ref{sec:data}. This relation was  calibrated in Paper I with the
help  of  a  set  of  high-resolution $N$-body  simulations  with  the
concordance    $\Lambda$    cold    dark   matter    cosmology.     In
Figure~\ref{fig:mhalo}, we plot the halo mass estimated in this way as
functions  of  both stellar  mass  (left  panel)  and central  stellar
velocity  dispersion (right  panel). Results  for the samples of different
$M_\ast$ and $\sigma_\ast$ are plotted in colorful symbols, with the 
size of the symbols being scaled by the halo mass. For  comparison, 
we have performed the same analysis
for a set of $M_\ast$  intervals without further dividing the galaxies
in each interval into subsamples of $\sigma_\ast$, as well as a set of
$\sigma_\ast$  intervals without  further dividing  the  galaxies into
subsamples of $M_\ast$.  This  gives an {\em average} relation between
halo mass and $M_\ast$, and  between halo mass and $\sigma_\ast$.  The
results are plotted  as solid triangles connected with  solid lines in
the figure.

The figure reveals two facts: a) that both $M_\ast$ and $\sigma_\ast$ 
are correlated with halo mass, and b) that the correlation between 
$M_\ast$ and halo mass is much tighter than the correlation between 
$\sigma_\ast$ and halo mass. The rms scatter of the different 
$\sigma_\ast$ samples around the average relation between halo mass and 
$M_\ast$ is 15.2\% or 0.06 dex, compared to 73.2\% or 0.24 dex
for the different $M_\ast$ samples around the average relation
between halo mass and $\sigma_\ast$. 

We finish this section by highlighting the stronger dependence of halo
mass  on $M_\ast$ than  on $\sigma_\ast$  in Figure~\ref{fig:samples},
where we indicate the halo mass of each sample by the size of a
red circle.

\section{Summary and discussion}

Using data from the SDSS DR7, we have derived  the velocity dispersion 
profiles for galaxy groups with central galaxies of different stellar  
masses ($M_\ast$) and stellar  velocity dispersions ($\sigma_\ast$). 
From these we have
obtained  estimates of  the  dark  matter halo  mass  for the  central
galaxies, and investigated the correlation of halo mass for central
galaxies with  $M_\ast$ and $\sigma_\ast$. At fixed  $M_\ast$, we find
very weak  or no correlation  between halo mass and  $\sigma_\ast$. In
contrast,   strong  mass   dependence  is   clearly  seen   even  when
$\sigma_\ast$ is limited to a narrow range.

\citet{Wake-Franx-vanDokkum-12}      recently     investigated     the
cross-correlation  functions between  galaxies of  given  $M_\ast$ and
$\sigma_\ast$ and  the parent  sample, finding $\sigma_\ast$  to be
more closely related than $M_\ast$ to the large-scale amplitude of the
correlation   functions.   The   authors  suggested   three   possible
explanations: 1) that halo mass for central galaxies
is more  closely related  to $\sigma_\ast$ than  to $M_\ast$,  2) that
halo  age (or  concentration)  for central  galaxies  is more  closely
related to $\sigma_\ast$ than to $M_\ast$, and 3) that halo properties
are still more  tightly related to $M_\ast$ than  to $\sigma_\ast$ and
the dependence  of clustering on  $\sigma_\ast$ at fixed  $M_\ast$ are
attributed to the contribution of satellite galaxies which deviate from
the stellar mass--halo mass relation of centrals.

Our  measurements of cross-correlation  functions between  galaxies and
central  galaxies and  velocity dispersion  profiles, as  well  as the
inferred halo  masses, are all  consistent with the  third possibility
being the  correct, or the most compelling  explanation.  As discussed
in  \citet{Wake-Franx-vanDokkum-12}, tidal  stripping  may reduce  the
size  and mass  of  a satellite  galaxy  as it  orbits  in its  parent
halo. This process has stronger effect on stellar mass than on central
stellar velocity  dispersion, and is  stronger in more  massive halos.
Therefore, at fixed stellar mass, galaxies of higher $\sigma_\ast$ are
more likely the  satellites in higher mass halos,  thus clustering more
strongly than  those of  lower $\sigma_\ast$. At  fixed $\sigma_\ast$,
galaxies of  higher $M_\ast$ are  more likely the satellites  in lower
mass halos, thus lowering  down the clustering amplitude and canceling
out the mass dependence to some extent.

As can  be seen from  the right-hand panel  of Figure~\ref{fig:mhalo},
the  stellar  velocity  dispersion   of  central  galaxies  is  indeed
correlated with halo  mass, as expected, but with  much larger scatter
when  compared to  stellar mass.  It is  clear that  the mass  of dark
matter  halos  is  correlated  more  tightly  with  the  stellar  mass
($M_\ast$)  of their  central galaxy,  than with  the  central stellar
velocity dispersion  ($\sigma_\ast$) of  the galaxy. Once  the stellar
mass  is   fixed,  the   central  velocity  dispersion   shows  little
correlation  with halo  mass.  Our  results thus  firmly rule  out the
other two possibilities proposed by \citet{Wake-Franx-vanDokkum-12}.

\acknowledgments

CL thanks David Wake for helpful discussion, and acknowledges the  
support of  the 100  Talents Program  of Chinese
Academy    of    Sciences    (CAS),   Shanghai    Pujiang    Programme
(no. 11PJ1411600) and the  exchange program between Max Planck Society
and  CAS.   This  work  is  sponsored  by  NSFC  (11173045,  11233005,
10878001,   11033006,  11121062)   and  the   CAS/SAFEA  International
Partnership Program for Creative Research Teams (KJCX2-YW-T23). This
work has made use of data from the SDSS and SDSS-II. The SDSS Web  Site 
is  http://www.sdss.org/.  


\label{lastpage}
\end{document}